\documentclass[aps,pra,twocolumn,groupedaddress, showpacs, showkeys]{revtex4-1}
\usepackage{braket}
\usepackage{amssymb}
\usepackage{amsmath}
\usepackage{graphicx}
\usepackage{float}
\begin{document}
\title{Deutschian closed timelike curves can create entanglement}
\author{Subhayan Roy Moulick}
\email[]{subhayan@acm.org}
\author{Prasanta K. Panigrahi}
\email[]{pprasanta@iiserkol.ac.in}
\affiliation{Indian Institute of Science Education and Research Kolkata, Mohanpur 741246, West Bengal, India}


\begin{abstract}
We study the nature of entanglement in presence of Deutschian closed timelike curves (D-CTCs) and we observe that qubits traveling along a D-CTC allow unambiguous discrimination of Bell states with Local Operations and Classical Communications (LOCC), that is otherwise known to be impossible. A consequence of this leads us to discover that localized D-CTCs can create entanglement between two parties, using just local operations and classical communication. This contradicts the fundamental definition of entanglement. 
\end{abstract}


\maketitle

Entanglement and Closed Timelike Curves(CTC) are perhaps the most exclusive features in quantum mechanics and general theory of relativity (GTR) respectively. Interestingly, both theories, advocate nonlocality through them.
While the existence of CTCs \cite{G49} is still debated upon, there is no reason for them, to not exist according to GTR \cite{MTY88, VW38}. CTCs come as a solution to Einstein's field equations, which is a classical theory itself. Seminal works due to Deutsch \cite{D91}, Lloyd et al. \cite{LMGGS11}, and Allen \cite{A14} have successfully ported these solutions into the framework of quantum mechanics. 
The formulation due to Lloyd et. al, through post-selected teleportation (P-CTCs) have been also experimentally verified \cite{Lothers11}.

The existence of CTCs has been disturbing to some physicists, due to the paradoxes, like the \emph{grandfather paradox} or the \emph{unproven theorem paradox}, that arise due to them.  
Deutsch resolved such paradoxes by presenting a method for finding self-consistent solutions of CTC interactions. The Deutschian model of CTCs (D-CTCs) impose a boundary condition, in which the density operator of the CTC system that interacts with a chronology respecting (CR) system is the same, both before and after it enters the wormhole. Formally, 
$$ \rho_{CTC} = \Phi(\rho_{CTC}) = Tr_{CR} \big( U (\rho_{CR} \otimes \rho_{CTC}) U^ {\dagger} \big) $$

where $\rho_{CR}$ is the density matrix for chronology-respecting system, $\rho_{CTC}$ is the initial density matrix of the qubit traveling along the closed timelike curve, and $U$ is the interaction unitary. Mathematically, this can be seen as nature finding a fixed point solution of the map, $\Phi$, that depends on the chronology respecting system \cite{D91}.

Although a complete theory of quantum gravity is yet to be formulated, quantum information theorists have been studying the implications of the existence of CTCs and the nature of information with CTC-assisted models of computation. Here, we turn our attention to understand the implications of existence of D-CTCs on entanglement. Recent studies of CTC-assisted models of computation, show them to be extremely powerful and be able to carry out non-trivial tasks, such as distinguish between non-orthogonal states \cite{BHW09, BW12}, clone unknown quantum states \cite{BWW13, AMRM13}, be able to signal superluminally \cite{BS14} and find a solution of any problem in the computational class PSPACE efficiently, in polynomial time (PSPACE=P) \cite{A09}.



We begin by understanding the problem of Bell state discrimination and ask if it might be possible to distinguish between Bell states, using only local operations and classical communication (LOCC), given only a single copy of the state, from a set of four Bell States. We then try to understand its implications. 

It is known, in the conventional model of quantum mechanics, it is possible to distinguish between any two Bell states using LOCC \cite{WSHV00}, however it is impossible to deterministically discriminate between four or even three Bell states \cite{GKRSS01}.  Here we take another look and study the problem of Bell state discrimination with the assumption of the existence of D-CTCs in nature. 

\emph{Bell state discrimination with LOCC} is defined as follows. Suppose a referee, prepares a single copy of a maximally entangled Bell state 
$$\Ket{\varphi}_{AB} \in_R \{ \Ket{\Phi^+}_{AB}, \Ket{\Phi^-}_{AB},  \Ket{\Psi^+}_{AB}, \Ket{\Psi^-}_{AB} \}$$
where $\Ket{\Phi^{\pm}} = \frac{1}{\sqrt{2}} \big(\Ket{00} \pm \Ket{11} \big)$ and $\Ket{\Psi^{\pm}} = \frac{1}{\sqrt{2}} \big(\Ket{01} \pm \Ket{10} \big)$, and gives one qubit to Alice ($\Ket{\varphi}_{A}$) and one qubit to Bob ($\Ket{\varphi}_{B}$), who are spatially separated and allowed only local operations and classical communication. Their (Alice and Bob's) objective is to determine which state was given to them.

One strategy Alice and Bob can pick would be the following. 

Alice prepares a (known) state $\Ket{\psi} = \alpha \Ket{0} + \beta \Ket{1}$, $0<\alpha \neq \beta<1$ and perform a Bell measurement on her (known) state and her part of the local entangled qubit, $\Ket{\varphi}_{A}$, and classically communicates the measurement outcomes to Bob. Depending on the Bell state Alice and Bob were sharing, the decomposition can be given as follows, for each of the four possible Bell states.

\begin{widetext}
\begin{center}
$\Ket{\psi}_{A} \Ket {\Phi^+}_{AB} = \frac{1}{2} \big( \Ket{\Phi^+}_A \Ket{\psi}_B + \Ket{\Phi^-}_A (Z\Ket{\psi}_B) + \Ket{\Psi^+}_A (X\Ket{\psi}_B) + \Ket{\Psi^-}_A (ZX\Ket{\psi}_B) \big)$\\
$\Ket{\psi}_{A} \Ket {\Phi^-}_{AB} = \frac{1}{2} \big( \Ket{\Phi^+}_A (Z\Ket{\psi}_B) + \Ket{\Phi^-}_A \Ket{\psi}_B + \Ket{\Psi^+}_A (ZX\Ket{\psi}_B) + \Ket{\Psi^-}_A (X\Ket{\psi}_B) \big)$\\
$\Ket{\psi}_{A} \Ket {\Psi^+}_{AB} = \frac{1}{2} \big( \Ket{\Phi^+}_A (X\Ket{\psi}_B) + \Ket{\Phi^-}_A (ZX\Ket{\psi}_B) + \Ket{\Psi^+}_A \Ket{\psi}_B + \Ket{\Psi^-}_A (Z\Ket{\psi}_B) \big)$\\
$\Ket{\psi}_{A} \Ket {\Psi^-}_{AB} = \frac{1}{2} \big( \Ket{\Phi^+}_A (ZX \Ket{\psi}_B) + \Ket{\Phi^-}_A (X\Ket{\psi}_B) + \Ket{\Psi^+}_A (Z\Ket{\psi}_B) + \Ket{\Psi^-}_A \Ket{\psi}_B \big)$
\end{center}
\end{widetext}

Bob performs the necessary unitary operations on his share of the entangled qubit depending on the classical communication from Alice, as follows, (I, X, Y, Z are the Pauli operators),
\begin{center}
$ 00 \rightarrow I, \hspace{1em} 01 \rightarrow X, \hspace{1em} 10 \rightarrow Z, \hspace{1em} 11 \rightarrow Y$   
\end{center}
In a sense, they `force'  the teleportated states to pick up the unitary error associated with Alice's Bell state measurement $\Ket{\Phi^+}_A$.
Now all that remains for Bob is to mark out the unitary error his resultant state contains.  
For this, Bob uses a variant of BHW-circuit \cite{BHW09} to distinguish between non-orthogonal states $\{ \alpha \Ket{0} + \beta \Ket{1} , \alpha \Ket{0} - \beta \Ket{1},\alpha \Ket{1} + \beta \Ket{0},\alpha \Ket{1} - \beta \Ket{0} \}$, as implemented in Fig.1, where the unitaries are defined as 

{\small
$
U_{00} = \begin{bmatrix} \alpha & \beta \\ -\beta & \alpha \end{bmatrix} \otimes \mathbb{I},  \hspace{2em}   
		U_{01} = (X \otimes X) \circ ( \begin{bmatrix} \beta & \alpha \\  \alpha & -\beta \end{bmatrix} \otimes \mathbb{I} ), $
		
$U_{10} =  (X \otimes \mathbb{I}) \circ ( \begin{bmatrix} \beta & \alpha \\ -\alpha & \beta \end{bmatrix} \otimes \mathbb{I}  ), \hspace{2em} 
		U_{11} = {\begin{bmatrix}  \alpha & \beta \\ \beta &  -\alpha  \end{bmatrix} \otimes X }
$
}%

The circuit first swaps the CTC system with the CR system. Following that it performs a controlled unitary with the CR systems as the control and CTC systems as the target. Finally, it measures the CR system in the computational basis. The CTC system is a nonlinear system. This is because the outcome of $\rho_{CTC}$, after the desired interactions, depends on the initial $\rho_{CTC}$ (before the interactions) and the CR system $\rho{CR}$. Also, $\rho_{CTC}$ (before the interactions) depends on CR system $\rho{CR}$. The objective here is to harness the non-linearity and exploit the two CTC qubits to effect the following map
\vspace{-0.5em}
\begin{center}
$(\alpha \Ket{0} + \beta \Ket{1}) \otimes \Ket{0} \rightarrow \Ket{00} ,$ \\
$(\alpha \Ket{0} - \beta \Ket{1}) \otimes \Ket{0} \rightarrow \Ket{01},$ \\
$(\beta \Ket{0} + \alpha \Ket{1}) \otimes \Ket{0} \rightarrow \Ket{10},$ \\
$(\beta \Ket{0} - \alpha \Ket{1}) \otimes \Ket{0} \rightarrow \Ket{11}.$ 
\end{center}
\vspace{-0.75em}
It can be seen, that these self-consistent solutions for the CTC qubits are unique, and satisfy Deutsch's criteria.

\begin{figure}[htbp]
\begin{center}
\includegraphics[scale=0.45]{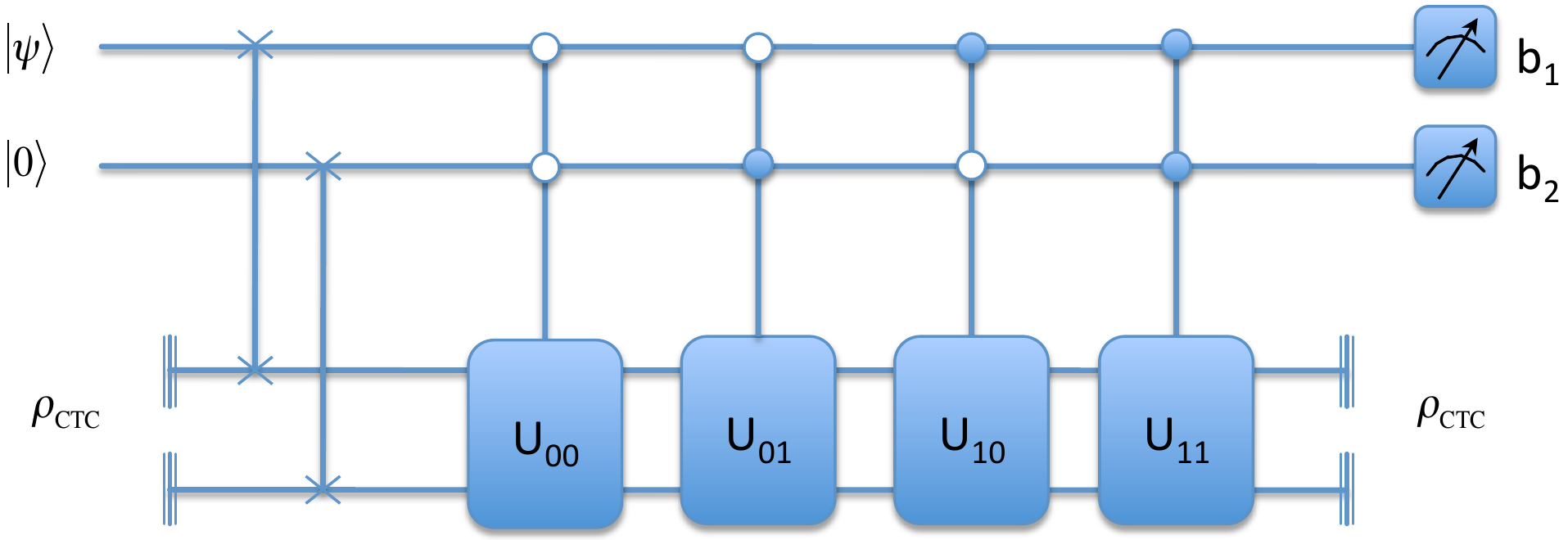}
\caption{BHW circuit to distinguish between states $\{ \alpha \Ket{0} \pm \beta \Ket{1}, \alpha \Ket{1} \pm \beta \Ket{0} \}$ using Deutschian formulations of CTC.}
\label{default}
\end{center}
\end{figure}

Let us understand one instance of what is happening in the circuit. Suppose the teleported state was $\Ket{\psi} = \alpha \Ket{1} + \beta \Ket{0}$. According to the desired interaction, it first swaps the information in the CR system and the CTC system. So, the CTC system now carries  $(\alpha \Ket{1} + \beta \Ket{0})\otimes \Ket{0}$. Since the CR system is now carrying $\Ket{1} \otimes \Ket{0}$, which the CTC system was initialized as, before the swap; unitary $U_{10}$ now acts on the CTC system and results in the CTC system to become $\Ket{1} \otimes \Ket{0}$, before it disappears in the wormwhole. Thus Deutsch's criteria for chronology respecting system is met and the qubits traveling along a CTC path remain the same both before and after the interaction. 
 
What is essentially happening here is Alice prepares a known state, $\Ket{\psi}$, and teleports it to Bob. The information of an entangled channel are not stored in the states but in the correlations. By teleporting the state, through the entangled channel, $\Ket{\psi}$ is affected by the correlation. In a sense, the correlation of the entanglement gets downloaded in the state. By studying the change of the teleported state from the prepared state, it becomes possible to understand the nature of correlation in the channel.
The circuit then, by measuring $b_1$ and $b_2$, of the chronology respecting qubits, learns which of the two conjugate eigenstates (through measurement $b_1$) and the eigenvalue ($(-1)^{b_2}$), the teleported state is in. 

The distinguishability of non-orthogonal states allows Bob to conclusively determine the Bell state that he shared with Alice. The corresponding Bell states, compared to the state identified by Bob, using the BHW circuit (Fig 1), are shown in Table 1. 

\begin{center}
Table 1:  Corresponding Bell States Alice \& Bob share\\
\begin{tabular}{|c|c|c|}
\hline 
{\bf Measurements}	&{\bf State Identified}		& {\bf Conclusive} \\ 
{\bf Outcomes, $b_1,b_2$}	&	{\bf by Bob}							& {\bf Bell State} \\  
\hline 
0,0	&	$\alpha \Ket{0} + \beta \Ket{1}$ 	&	$\Ket{\Phi^+}$ \\
\hline 
0,1	&	$\alpha \Ket{0} - \beta \Ket{1}$ 		&	$\Ket{\Phi^-}$ \\ 
\hline 
1,0	&	$\alpha \Ket{1} + \beta \Ket{0}$ 	&	$\Ket{\Psi^+}$ \\  
\hline 
1,1	&	$\alpha \Ket{1} - \beta \Ket{0}$ 		&	$\Ket{\Psi^-}$ \\ 
\hline
\end{tabular}
\end{center}

This strategy works efficiently to discriminate between the Bell states, if we could use D-CTCs. 
But what does this say about entanglement in general? Consider the Smolin State, a certain four-party unlockable bound-entangled state \cite{S01}, shared between Alice, Bob, Charlie \& Dan,   
\begin{widetext}
{
\begin{eqnarray*}
\rho =  & \frac{1}{4}( \Ket{\Phi^+}\Bra{\Phi^+}^{AB} \otimes \Ket{\Phi^+}\Bra{\Phi^+}^{CD} + \Ket{\Phi^-}\Bra{\Phi^-}^{AB} \otimes \Ket{\Phi^-}\Bra{\Phi^-}^{CD} +	\\ 		
		&			 \Ket{\Psi^+}\Bra{\Psi^+}^{AB} \otimes \Ket{\Psi^+}\Bra{\Psi^+}^{CD} + \Ket{\Psi^-}\Bra{\Psi^-}^{AB} \otimes \Ket{\Psi^-}\Bra{\Psi^-}^{CD} 
 )
 \end{eqnarray*} 
 }%
\end{widetext}

It can be seen that entanglement between AB and CD is 0, i.e
	$$\varepsilon(AB:CD) = 0$$
and the state is invariant under permutation. Thus, 
	$$\varepsilon(AB:CD) = \varepsilon(AC:BD) = \varepsilon(AD:BC) = 0$$
In other words, $\rho$ is separable across the three bipartite cuts AB : CD, AC : BD and AD : BC \cite{S01}. 

The logarithmic negativity \cite{VW02}, $E_N$, of the state $\rho$, in AC:BD cut, is  
	$$E_N(\rho) = log_2 || (\rho^{T})^{AC} ||_1  = log_2 ( |1/{\sqrt{2}}|^2 + |1/{\sqrt{2}}|^2) = 0$$

Since the distillable entanglement, $E_D$, is upper bounded by logarithmic negativity \cite{VW02}, we can say
	$$ E_D(\rho) \leq E_N(\rho) =0$$
Thus distillable entanglement is exactly zero for a Smolin state. 

Now since CTC-assisted computation allow discrimination allow Bell State discrimination, given the Smolin state to Alice, Bob, Charlie and Dan, Alice and Bob can distinguish their Bell state without meeting. Following that, Alice and Bob classically communicate their Bell states to Charlie and Dan respectively, who now have share a maximally entangled Bell state. Hence $1-ebit$ was distilled using only local operations and classical communication, from the Smolin state through a D-CTC assisted computation. This shows, existence of D-CTCs would imply the possibility of creating entanglement using LOCC, which is otherwise impossible according to current formulation of quantum mechanics.

To conclude, our work here raises fundamental questions concerning the nature of entanglement in a  world with Deutschian closed timelike curves, that drastically changes our current understanding of quantum mechanics. An intuitive resolution to this might lead to support the chronology protection conjecture \cite{H92}, which loosely says such closed timelike curves cannot exist in nature. If this were to be indeed true, such contradictions could indeed be evaded. However, it was recently shown that one can also replicate the effects of Deutschian closed timelike curves in quantum states, in chronology respecting open timelike curves \cite{Yother15}. So, in a sense, this may not be a problem with the Deutschian formalism, but a problem in nature.
A full theory of quantum gravity, we expect would perhaps resolve such challenges and contradiction between the implications of CTCs and laws of quantum mechanics and hope this work will help motivate further research.

\bibliography{references}

\end{document}